# Systematics of nucleon density distributions and neutron skin of nuclei

W. M. Seif and Hesham Mansour

Physics Department, Faculty of Science, Cairo University, Egypt

**Abstract**

Proton and neutron density profiles of 760 nuclei in the mass region of $A = 16 - 304$ are analyzed using the Skyrme energy density for the parameter set SLy4. Simple formulae are obtained to fit the resulting radii and diffuseness data. These formulae are useful to estimate the values of the unmeasured radii, and especially in extrapolating charge radii values for nuclei which are far from the valley of stability or to perform analytic calculations for bound and/or scattering problems. The obtained neutron and proton root-mean-square radii and the neutron skin thicknesses are in agreement with the available experimental data and previous Hartree-Fock calculations.

## I. Introduction

To get accurate information on the density distributions of finite nuclei is of high priority in nuclear physics. Basically, to describe perfectly finite nuclei we have to start with full information regarding the root-mean-square (rms) radii of their proton and neutron density distributions, their surface diffuseness, and their neutron skin thickness. Unlike the available experimental data on the charged proton distributions, the available data for the neutral neutron distributions and their rms radii inside nuclei, as well as the neutron skin thickness are not enough yet. Previous extensive studies are aimed to investigate the correlation between the differences in the proton and neutron radii in finite nuclear systems and the nuclear symmetry energy [1, 2, 3, and 4] and neutron skin thickness. Most works are restricted to spherical nuclei. The accurate investigation of the nuclear density distributions is of special importance to explore the saturation properties of asymmetric nuclear matter and its equation of state [5]. On the other hand, the clear correlation between the finite surface diffuseness and the number of surface nucleons influences the surface and the outer region of the nucleus-nucleus interaction potential [6] as well as the clustering of the individual entities on the nuclear surface. This would definitely affect the different nuclear reactions and the nuclear decay processes, as well as different astrophysical quantities. Also, the value of the surface diffuseness affects the position of the Coulomb barrier between any interacting nuclei [7]. The experimental data on the density distributions of neutrons inside stable and exotic nuclei are mainly focussing on measuring their rms radii [8, 9] and the neutron skin thickness [10]. Protons radii and moments of their distributions in finite nuclei are known from the charge probing experiments [8] and the ones related to the charge-dependent nuclear properties. The corresponding experiments on the neutron distributions give some little information about their higher moments [8,9] and deformations. These distributions are normally form a "neutron skin". The neutron skin thickness is typically defined as the difference between the rms radii of the neutron and proton density distributions in the nucleus. It is related to the differences between the equation of state of asymmetric nuclear matter and that of pure neutron matter. It is also related to the nuclear symmetry energy and its density dependence.

The neutron skin can also be used to probe the pressure of asymmetric nuclear matter and its influences to the pressure between neutrons and protons in finite nuclei. Several experiments can be used to study the neutron density distributions of atomic nuclei and their rms radii. For instance, the elastic scattering processes of proton [9,11,12, and 13] and $\alpha$-particle projectiles [14], and the experiments involving giant dipole and spin-dipole resonances [15,16]. Also, the radiochemical and x-ray data of antiprotonic atoms, such as the atomic level shifts and level widths, are usually proposed to this aim [17, 18,19,20,21, 22]. The precise analysis of such experimental data is of special interest not only to determine the density distribution and the neutron skin but also to investigate the different conditions influencing their bulk and surface parts. However, the reliable determination of the density dependence of the symmetry energy is also interesting to study the neutron density in atomic nuclei and asymmetric nuclear matter. This would affect the calculation of the heavy ion reactions and other investigations related to atomic parity violation. The advanced technique of exotic beams in modern accelerators as well as recent astrophysical observables raised the interest in the symmetry energy and neutron skin studies for stable and exotic nuclei. In the frame work of mean-field calculations, its noticed that the neutron skin thickness in neutron rich nuclei such as $^{208}$Pb gives a linear correlation with the slope of the equation of state of neutron matter at a neutron density of 0.10 fm$^{-3}$ [23, 24].

In the present work we theoretically investigate the neutron skin thickness of nuclei by calculating parameterized nucleon densities in a self-consistent way as will be shown in the next section. The purpose of the present work is to make systematic studies for a large number of nuclei with the least amount of numerical work. The energy density functional based on the Skyrme SLy4 effective interaction [25] is a suitable tool for such a study [26, 27, 28, 29,30,31]. Similar works have been reported in the literature using other approaches and approximations [28, 32, 33, 34, and 35]. The nuclear energy density functional is optimized to provide an adequate approximation of the total energy and the local nuclear local. The SLy4 parameterization of the Skyrme-like force is widely used in different nuclear studies such as the nuclear structure [36,37,38], the nuclear reactions [39,40,41] as well as the decay processes of heavy nuclei [42,43,44]. One of the advantages of the Skyrme force is that it gives simple analytic expressions which are easy to handle with straight forward calculations. Here, we make a large-scale analysis of the proton and neutron nuclear density profiles as well as the thicknesses along the whole nuclear chart. In the next section we present the outlines of the method of calculation. In section III we discuss the results obtained. Section IV is devoted for the summary and conclusions.

## II. **Formalism**:

In the present work we will consider a self-consistent Skyrme Hartree-Fock (HF) approach. The total energy of a nuclear system is given in terms of the local nucleon ($\rho_{n,p}$), kinetic ($\tau_{n,p}$), and spin-orbit ($\vec{J}_{n,p}$) densities [34]. These densities are in turn defined in terms of the single-particle wave functions ($\varphi^i(\sigma)$) and the occupation numbers ($n^i$). Here, $i$ and $\sigma$ are the orbital and spin quantum numbers, respectively. In this case, the local proton (neutron) density is obtained as a sum over the single-particle occupied states [25, 34],

$$\rho_{p(n)}(\vec{r}) = \sum_{i,\sigma} |\varphi^i_{p(n)}(\vec{r},\sigma)|^2 n^i_{p(n)}. \qquad (1)$$

The Skyrme energy density functional consists of a kinetic, nuclear (Skyrme) and Coulomb parts.

$$H(\rho_{n,p}, \tau_{n,p}, \vec{J}_{n,p}) = \frac{\hbar^2}{2m} \sum_{i=n,p} \tau_{n,p} + H_{Sky} + H_{Coul} \qquad (2)$$

The effective Skyrme-like interaction ($H_{Sky}$) contains zero- and finite-range, density-dependent, and effective-mass terms. Also, it contains a spin-orbit and tensor coupling with the spin and gradient terms included [25]. The energy density functional based on the Sly4 parameterization of the Skyrme effective interaction describe successfully the ground-state properties of finite nuclei [25]. We performed the HF calculations using the Skyrme-Hartree-Fock program of P.G. Reinhard [30] based on the Sly4 parameterization of the Skyrme interaction. For a practical use in the nuclear structure and reaction calculation, one would express the neutron and proton densities of the deformed nuclei in the two-parameter Fermi shape as,

$$\rho_{n(p)}(r,\theta) = \rho_{0n(p)}\left(1 + e^{(r-R_{n(p)}(\theta))/a_{n(p)}}\right)^{-1}. \qquad (3)$$

Here, $R_{0n(p)}$ and $a_{n(p)}$ are the half-density radius and the diffuseness of the neutron (proton) density distributions, respectively. The Fermi shape given by Eq. (3) normally allows an easy inclusion of the proton and neutron densities, and their derivatives, in any complicated calculations. In the present work, we shall deduce the half-density radius and diffuseness for the proton and neutron distributions of each nucleus from a fit to its density distributions obtained from the self-consistent Skyrme HF calculations. To get the accurate half-density radius and diffuseness and to exclude the fluctuations in the internal region of the calculated density, we fitted the obtained numerical density to the function $r^2\rho$ instead of the function $\rho$ which is given by the Fermi shape. The density parameters $\rho_{0n(p)}$ can be obtained from the normalization of the neutron (proton) density distributions to the total number of neutrons (protons) in the nucleus. The proton and neutron root mean square radii can be defined as,

$$R^{rms}_{n(p)} = \langle r^2_{n(p)} \rangle^{1/2} = \left(\frac{\int r^2_{n(p)} \rho_{n(p)}(\vec{r}) d\vec{r}}{\int \rho_{n(p)}(\vec{r}) d\vec{r}}\right)^{1/2} \approx \left(\frac{3}{5} R^2_{0n(p)} + \frac{7\pi^2}{5} a^2_{n(p)}\right)^{\frac{1}{2}} \qquad (4)$$

In this expression, terms of the order $of\ e^{-R_{0n(p)}/a_{n(p)}}$ and $(a/R)^4$ have been neglected [45]. For deformed nuclei, the deformed half-density radius becomes,

$$R_{n(p)}(\theta) = R_{0n(p)}[1 + \beta_2 Y_{20}(\theta) + \cdots]. \qquad (5)$$

where $\beta_2$ represents the quadrupole deformation. Neglecting the higher order deformation components and keeping only the leading terms up to the quadratic one, one can expand the deformed density distribution with respect to the deformation parameter $\beta_2$ to obtain [45],

$$R^{rms}_{n(p)} = \langle R^2_{n(p)} \rangle^{1/2} \approx \left[\left(\frac{3}{5} R^2_{0n(p)} + \frac{7\pi^2}{5} a^2_{n(p)}\right)\left(1 + \frac{5}{4\pi} \beta_2^2\right)\right]^{\frac{1}{2}} \qquad (6)$$

The neutron skin thickness which represents the extension of the neutron density can then be obtained as

$$\Delta_n = \langle R_n^2 \rangle^{1/2} - \langle R_p^2 \rangle^{1/2} \qquad (7)$$

## III. Results and discussion:

Fig. 1(a) shows the half-density radii of the neutron density distributions for the isotopic chains of O(Z=8), Mg(Z=12), Ar(Z=18), Cr(Z=24), Cu(Z=29), Se(Z=34), Zr(Z=40), Sn(Z=50), Pm(Z=61), Os(Z=76), Hg(Z=80), Th(Z=90), No(Z=102) and Fl (Z=114). The half-density radii of the neutron density distributions for the studied 760 nuclei in the mass region of $A = 16 - 304$ are presented in Fig. 1(b). Figs. 1(a) and 1(b), show clearly the general $N^{1/3}$ behaviour of the neutron half-radii. We notice that the increasing rate of the neutrons radius along the same isotopic chain is related to the shell and sub-shell closures of the associated protons. For example, the small slopes between brackets for the isotopic chains presented in Fig. 1(a) are for Z=8 (0.904), Z=50 (0.917), Z=80(0.947), Z=102(0.959), and Z=114 (0.697).This is the case for shell or sub-shell closures, or very close to a shell closure. The large slopes appear for Z=12 (1.113), Z=24 (1.020), Z=29(1.051), Z=61(1.028), and Z=90(1.087).Here, again the case which represent a partially occupied shells open shells. According to Fig.1 (b), one can find that the half-radius of the neutron distribution for a nucleus (Z, N) can be represented as

$$R_{0n} = 0.953\, N^{1/3} + 0.015\, Z + 0.774 \qquad (8)$$

We obtained this relation by fitting the results of all nuclei presented in Fig. 1(b). For the studied 760 nuclei, the standard deviation, $\sigma = \sqrt{\sum_{i=1}^{n}\left[log_{10}\left(X_i^{approx}/X_i^{HF}\right)\right]^2/(n-1)}$, of the calculated half-radii based on this empirical relation with respect to the exact values from the full HF calculations is $\sigma = 0.0068$. Also, an average error, $\chi(\%) = \frac{1}{n}\sum_{i=1}^{n}\left|\frac{X_i^{approx}-X_i^{HF}}{X_i^{HF}}\right| \times 100$, with a value of $\chi = 1.059\%$ is obtained. Fig. 2(a) shows the variation of the diffuseness of the neutron density distributions for the same isotopic chains which are shown in Fig. 1(a). The extracted diffuseness values of all studied nuclei are shown in Fig. 2(b). For the same isotopic chain, an oscillatory behaviour is observed in

Fig. 2(a) for the diffuseness values versus the neutrons number. This diffuseness shows a minimum value when the neutrons number approaches a closed or semi-closed shells. The diffuseness has a maximum value when the shells are half-occupied. For example, the diffuseness values for the isotopic chains of Sn are in the range between *a*=0.497 and *a*=0.657. One can also observe that two maximum diffuseness values are obtained at N=50 and N=82, as well as two minimum values are obtained at N=70 and N=98. Also, for the isotopic chains of Zr, Hg, and Th, the minimum diffuseness values are obtained around N=50, N=104 and N=164, respectively. The maximum diffuseness values of the same isotopic chains are obtained around N=70, N=140 and N=136, respectively. Fig. 2(b) shows the behaviour of the diffuseness with the values of N/Z. The diffuseness of the neutron distribution for a nucleus (Z,N ) is given by the following fitting equation:

$$a_n = 0.446 + 0.072 \left(\frac{N}{Z}\right) \qquad (9)$$

For the studied 760 nuclei, this empirical relation yields, the standard deviation $\sigma = 0.0177$ and an average error $\chi = 3.305\%$. Figs. 3(a) and 4(a), show respectively, the half-density radii and diffuseness of the proton density distributions for the isotonic chains of N= 8, 12, 18, 29, 40, 50, 61, 70, 82, 92, 104, 126, 140, 150, 170 and 184. The values of the deuced radii and diffuseness for all studied nuclei are plotted in Fig. 3(b), versus $Z^{1/3}$, and Fig. 4(b), versus Z/N, respectively. Again, the increasing rate of the protons radii along the same isotonic chain varies according to the number of neutrons in the open shells. The small slopes between brackets for the isotonic chains presented in Fig. 3(a) are for N=8 (0.907), N=50(1.011), N=82(0.990), N=104(0.974), N=126(1.024), N=150(0.995), N=170(0.919), and N=184(0.87).This case represents a shell or sub-shell closures, or very close to them. The large slopes appear for N=40 (1.055), N=61(1.085), N=70(1.031), and N=140(1.083). This would represent a mid-shell or open shells occupied with odd neutron number. In Fig. 3(b), one can fit the obtained half-radii of the presented proton distribution as

$$R_{0p} = 1.322\, Z^{1/3} + 0.007\, N + 0.022 \qquad (10)$$

For the studied nuclei, a standard deviation of $\sigma=0.0057$ and an average error of $\chi =0.824\%$ are obtained. Also, for the same isotonic chain, the same oscillatory behaviour of the diffuseness of the neutron distributions is obtained for the proton distributions versus the proton number. This diffuseness shows minimum values when the protons number approaches a closed or semi-closed shells of the protons. For example, the diffuseness values of the proton distributions of the isotonic chain of N=50 (N=92) are found to be ranging between $a=0.490$ (0.496) and $a=0.521$ (0.526). The minimum diffuseness values for the two chains of N=50 and N=92 are obtained around N=50. The maximum diffuseness values are obtained around N=40 and N=70, respectively. The fitting procedure for the proton density distributions of the studied nuclei, Fig. 4(b), yields

$$a_p = 0.449 + 0.071 \left(\frac{Z}{N}\right) \qquad (11)$$

According to this empirical relation, a standard deviation of $\sigma=0.0104$ and an average error of $\chi =1.881\%$ are obtained. Fig. 5 shows our calculated neutron skin thicknesses for Ca (Z=20), Fe, Zr, Sn, Te, Yb, Pb, Th, U, and Fm (Z=100). The experimental points were shown in the mentioned figure for $^{40,\,48}$Ca (Fig. 5(a)), $^{54,56,57}$Fe and $^{208}$Pb (Fig. 5(b)), $^{90,\,96}$Zr and $^{232}$Th (Fig. 5(c)), $^{112,114,116,118,120,122,124}$Sn and $^{238}$U (Fig. 5(d)), and $^{122,124,126,128,130}$Te (Fig. 5(e)). The calculated neutron skin thickness for the different presented isotopes, which is based on the simple empirical formulae, Eqs. (8, 9, 10, and 11), is in good agreement with the ones from the full HF calculations. The deviation starts to appear for the isotopes of high isospin asymmetry. Also, the deviation increases for the heavy isotopic chains of large Z. This is because one might be in need for a slightly different parameterization for the neutron and proton density distributions for the super-heavy nuclei. Fig. 5 also shows that most of the presented experimental data of the neutron skins are successfully evaluated by both our approximate formulae and the full HF calculations (15 experimental points out of 22). One can also notice that for the heavy nuclei of large Z, the uncertainty presented by the experimental error in their experimentally evaluated skin thickness increases. Finally, it is of interest to mention that the deformation effect on the values of the rms radii is noticeable whereas minor contribution is observed in the calculated values of the neutron skin thicknesses. For example due to its quadrupole deformation, the rms radii of the neutron and proton densities of $^{238}$U increases from 5.933 and 5.733 ($\beta_2=0$) to 5.988 and 5.786 ($\beta_2=0.215$), respectively. Consequently, its neutron skin thickness increases from 0.200 to 0.202 only. Perhaps, this is because of the inherent subtraction in Eq. (7).

## IV. **Summery and conclusion**:

In the present work we systemically studied the neutron and proton density distributions for a large number of nuclei (760). Our calculations are performed with some representative effective nuclear forces. That is the zero-range Skyrme SLy4 force from the nonrelativistic framework. The proton and neutron densities have been assumed to have a two-parameter Fermi shape and their parameters were obtained in a self consistent way. A minimization was required for the two different quantities standard deviation and average error; at the same time. An overall linear increasing behaviour of the radii of the neutron and proton density with the increase of $N^{1/3}$ and $Z^{1/3}$, respectively, is observed. Slight change of the slopes of these linear variations for the different isotopic or isotonic groups is also observed. An overall slight increase in the diffuseness of the neutrons and protons is observed with the increase of N/Z and Z/N, respectively. We also observed a clear oscillatory behaviour in the diffuseness with the number of nucleons in the open shells, along the same isotopic and isotonic chain. The diffuseness is found to have a minimum values for the closed shells nuclei. It increases with increasing the number of nucleons in the open shells. It shows also maximum values around the mid-shell occupations. It is also shown that the behaviour of the neutron skin thickness for different isospin values is in good agreement with the full HF calculations and the available experimental data. We investigated in the present work many new isotopes in the spirit as we mentioned of previous published works but with the statistics are noticeably enlarged. Semi empirical formulae were presented for the general behaviour of the radii and diffuseness of the nucleon densities. Of course more elaborate experimental data are needed. Also one can use extended Fermi distributions or more elaborate density distributions for a thorough investigation of the super-heavy deformed nuclei.

**Figures captions**

**Fig. 1**: The half-density radii of the neutron density distributions for (a) several isotopic chains (b) 760 nuclei in the mass region of $A = 16 - 304$.

**Fig. 2**: The diffuseness of the neutron density distributions for (a) several isotopic chains (b) 760 nuclei in the mass region of $A = 16 - 304$.

**Fig. 3**: The half-density radii of the proton density distributions for (a) several isotonic chains (b) 760 nuclei in the mass region of $A = 16 - 304$.

**Fig. 4**: The diffuseness of the proton density distributions for (a) several isotonic chains (b) 760 nuclei in the mass region of $A = 16 - 304$.

**Fig. 5**: The neutron skin thickness for several isotopic chains calculated from the approximated root-mean-square radii given by Eqs. (8) - (11). The figure also shows a comparison with the values deduced from the full HF calculations and the available experimental data obtained from the antiprotonic atom x-ray data [20, 21] and those from the ($^3$He, *t*) charge-exchange reaction [15] and coherent pion photo production [46] for some isotopes.

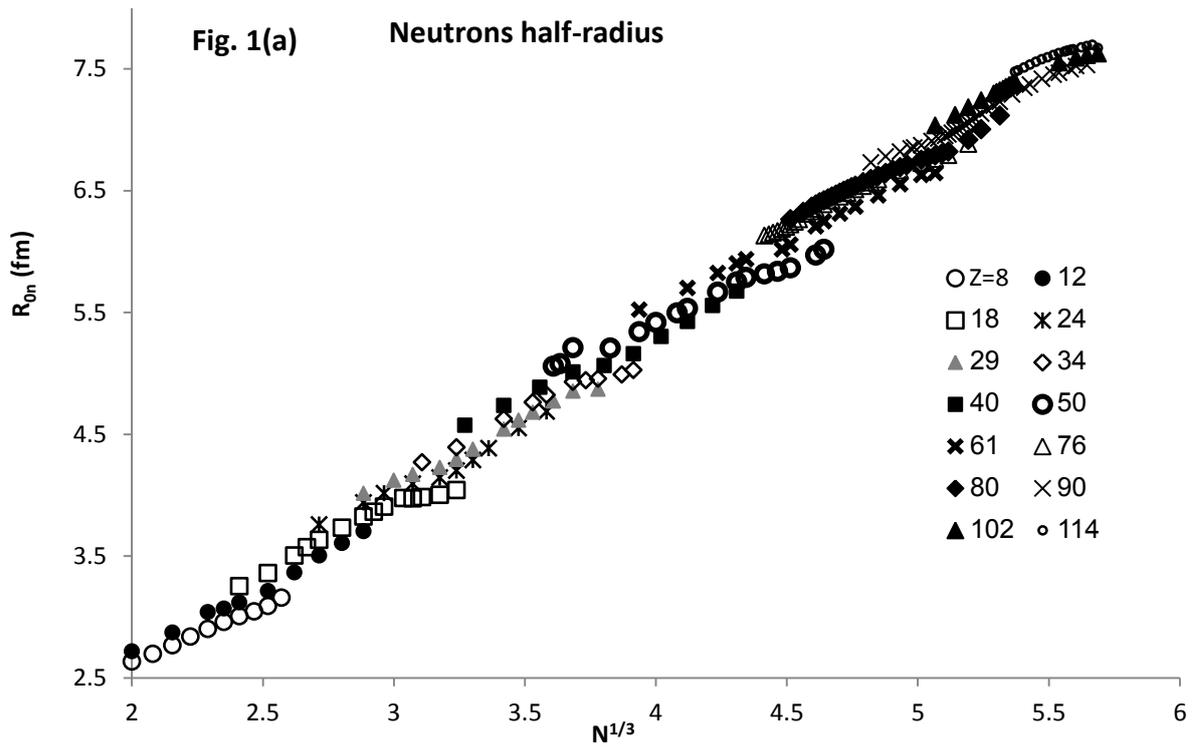

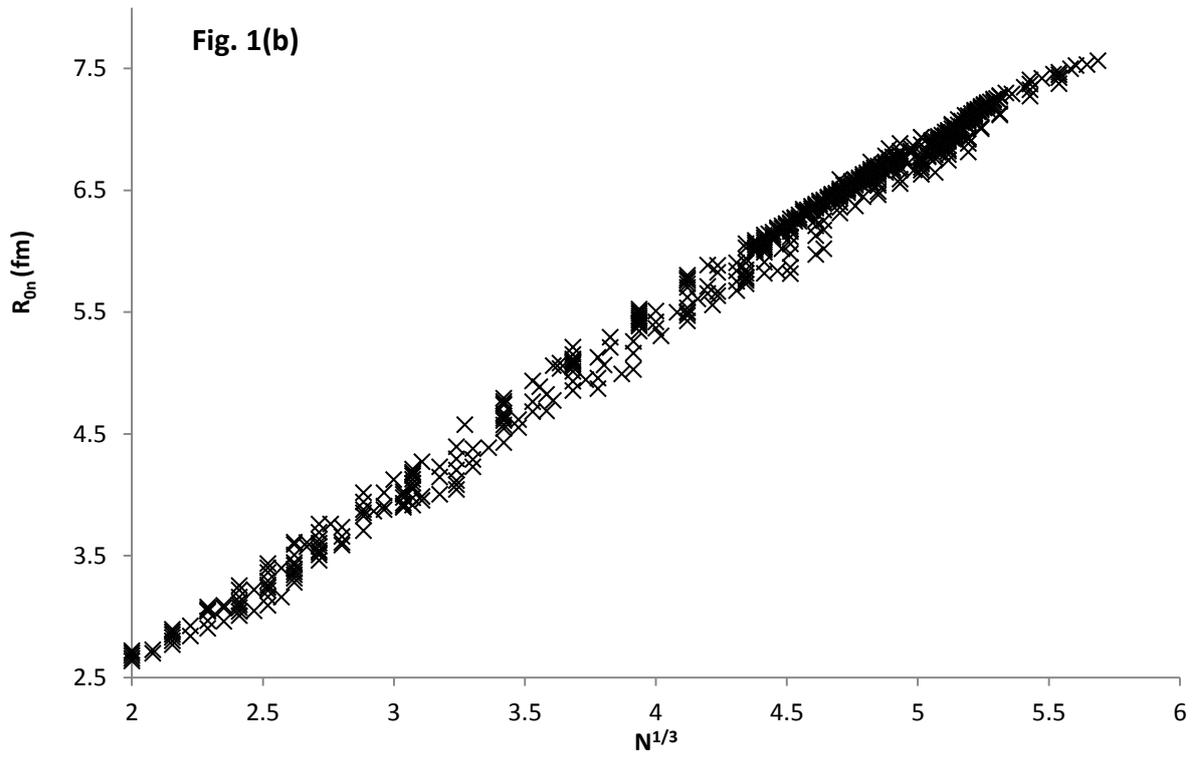

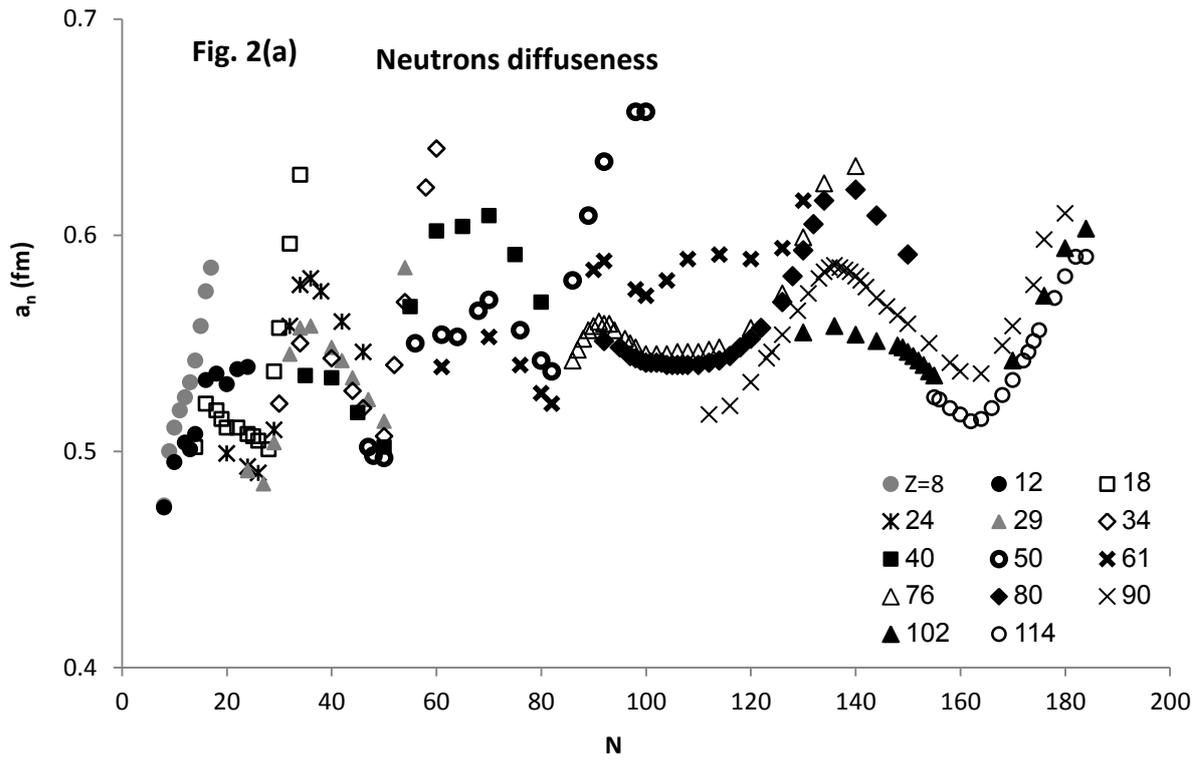

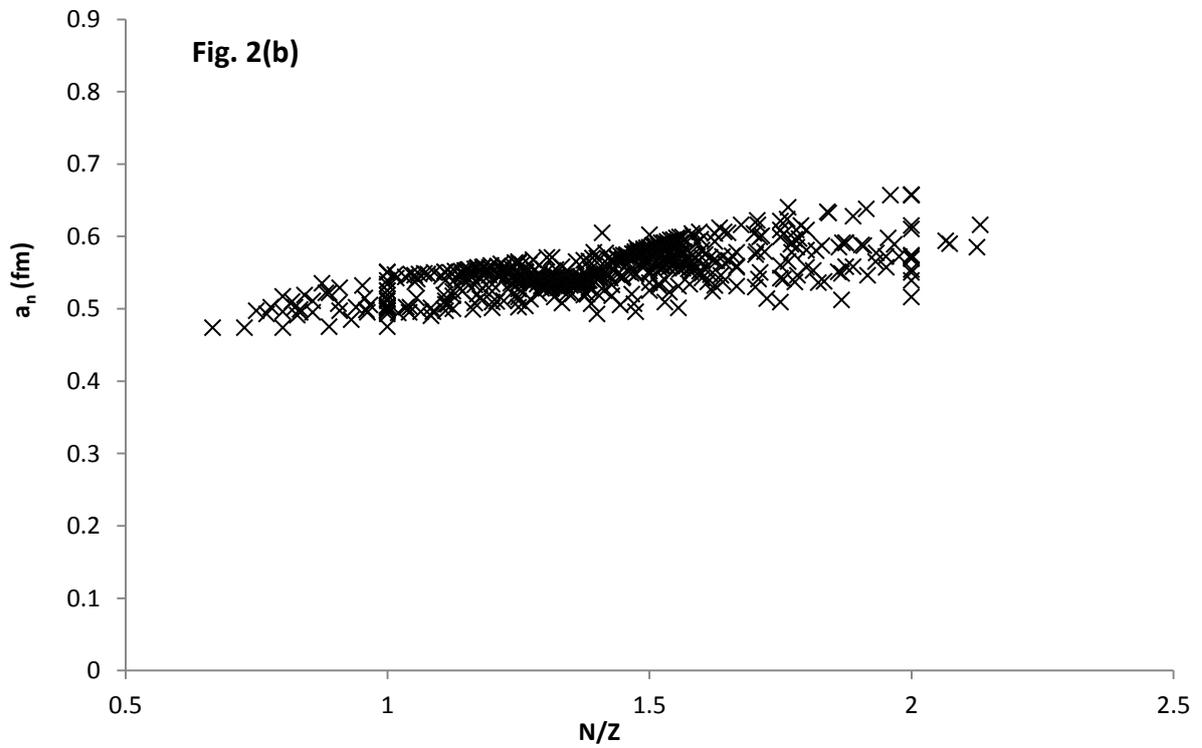

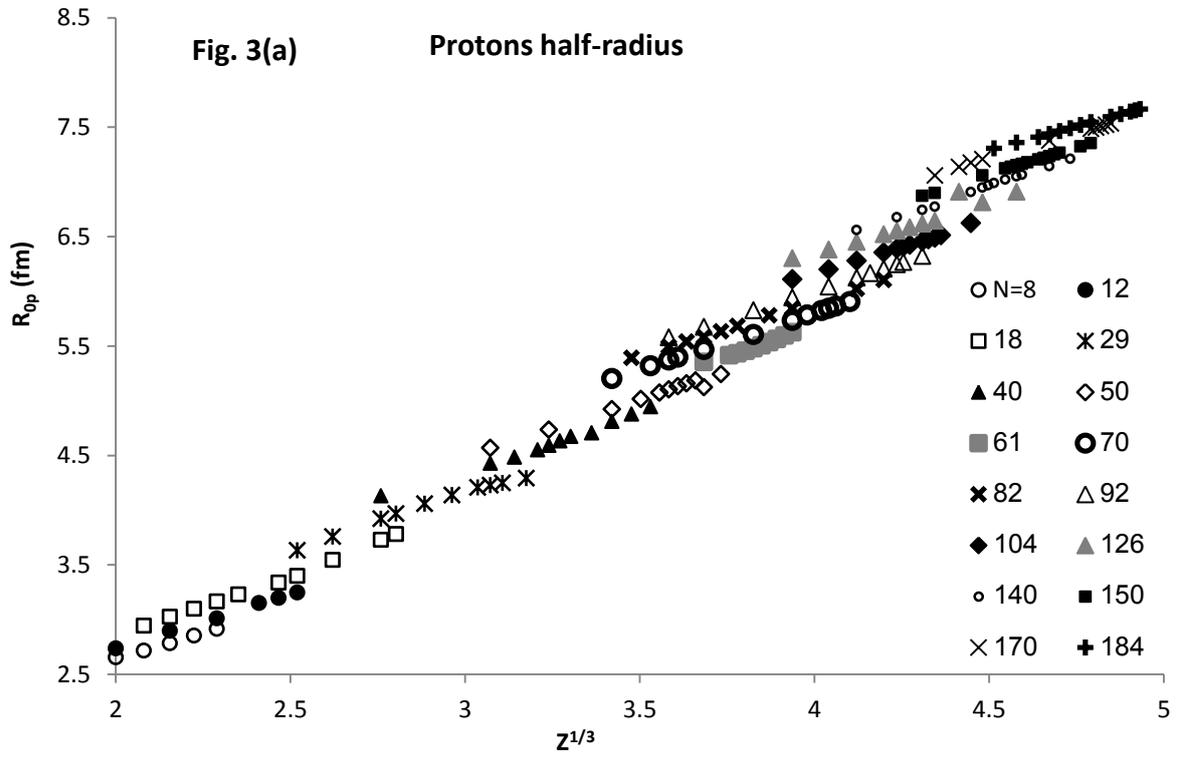

Fig. 3(a) Protons half-radius

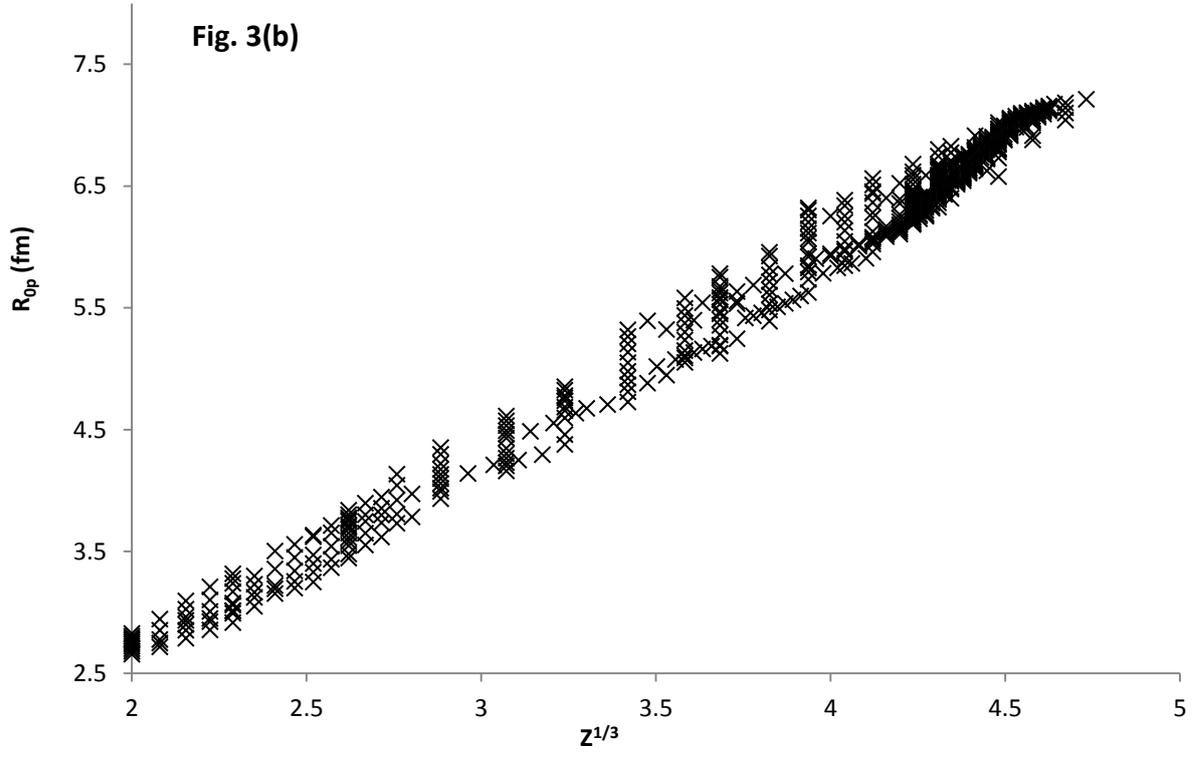

Fig. 3(b)

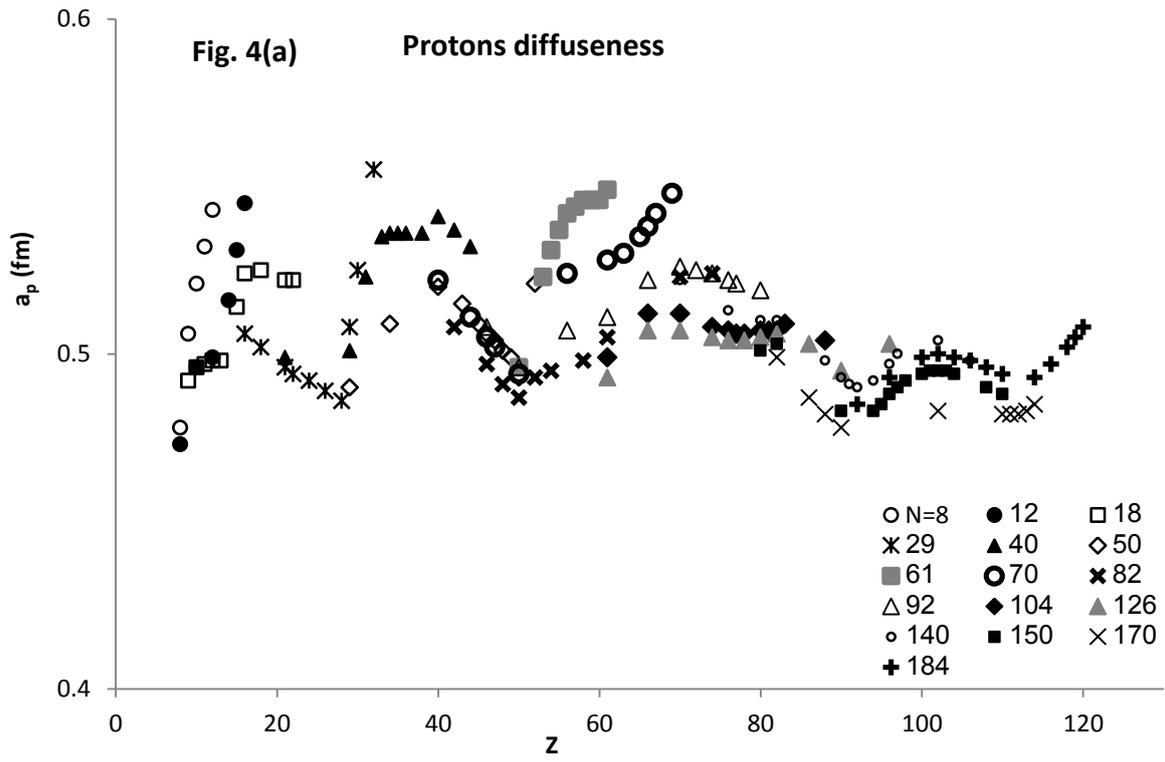

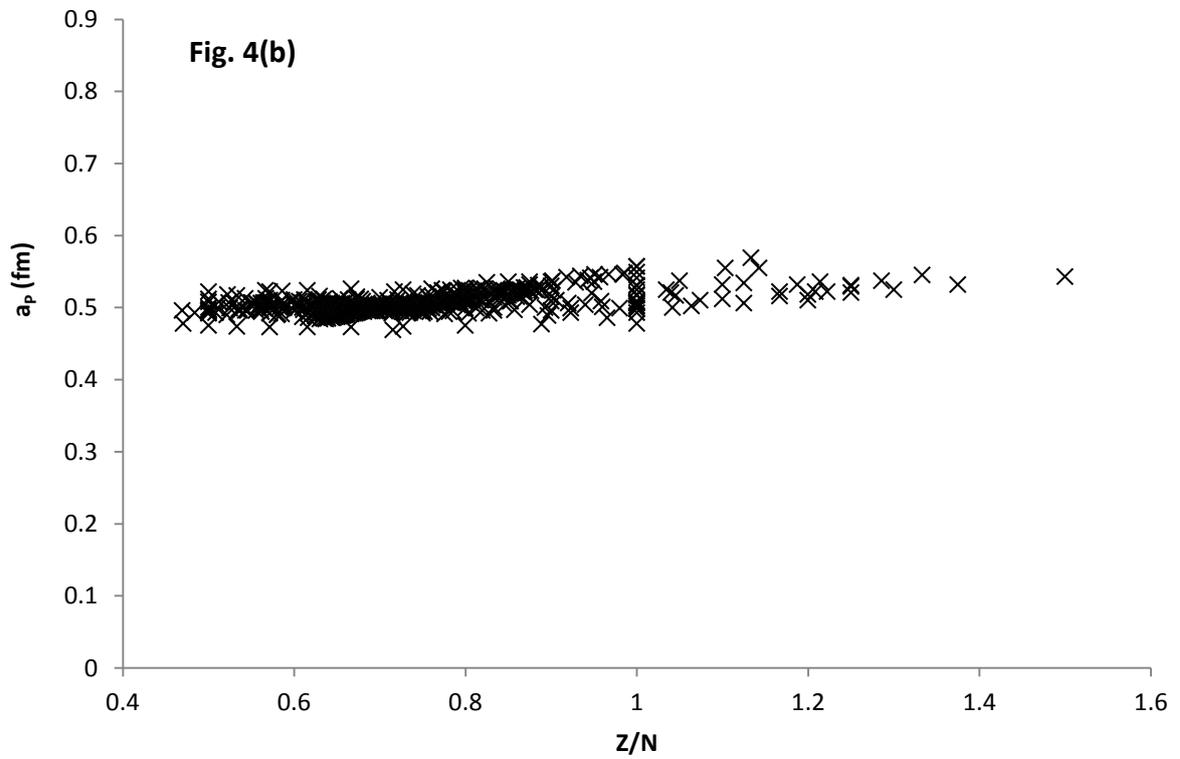

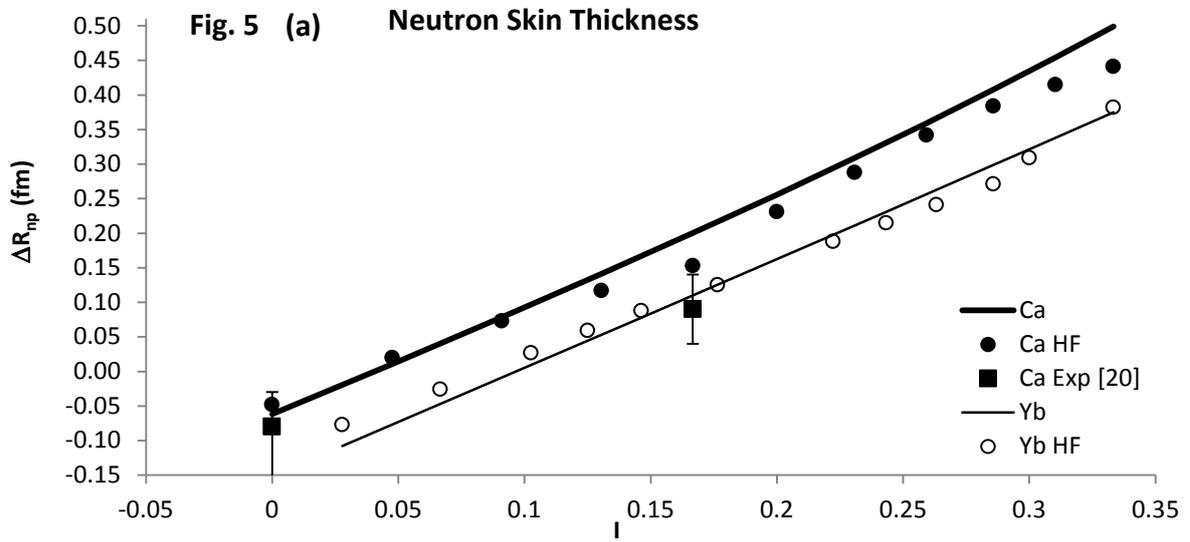

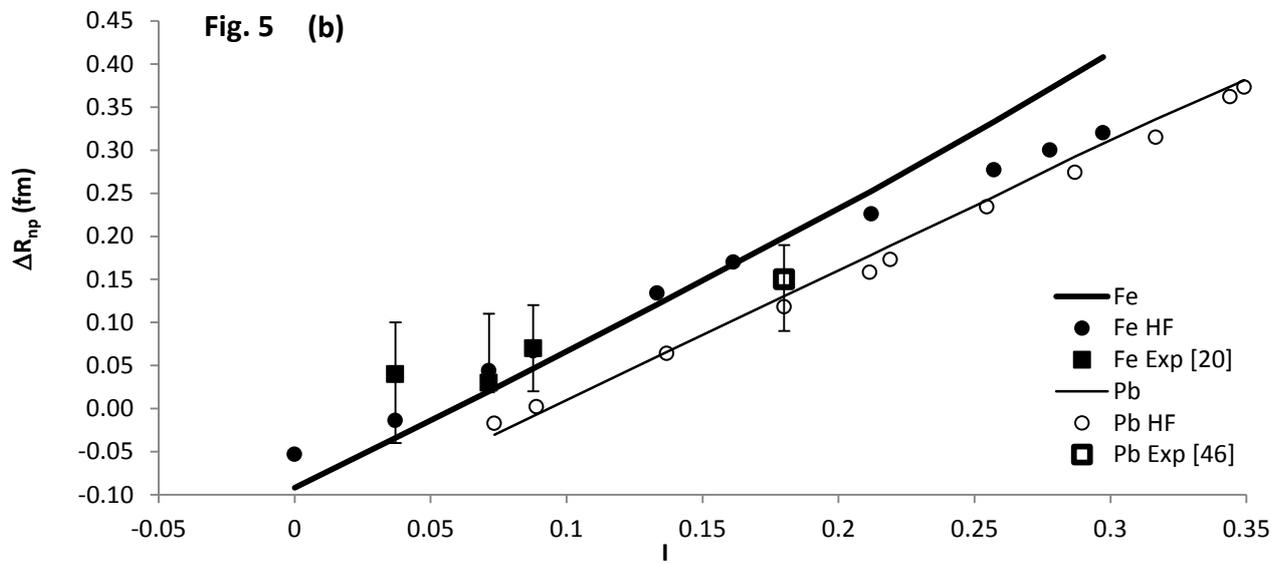

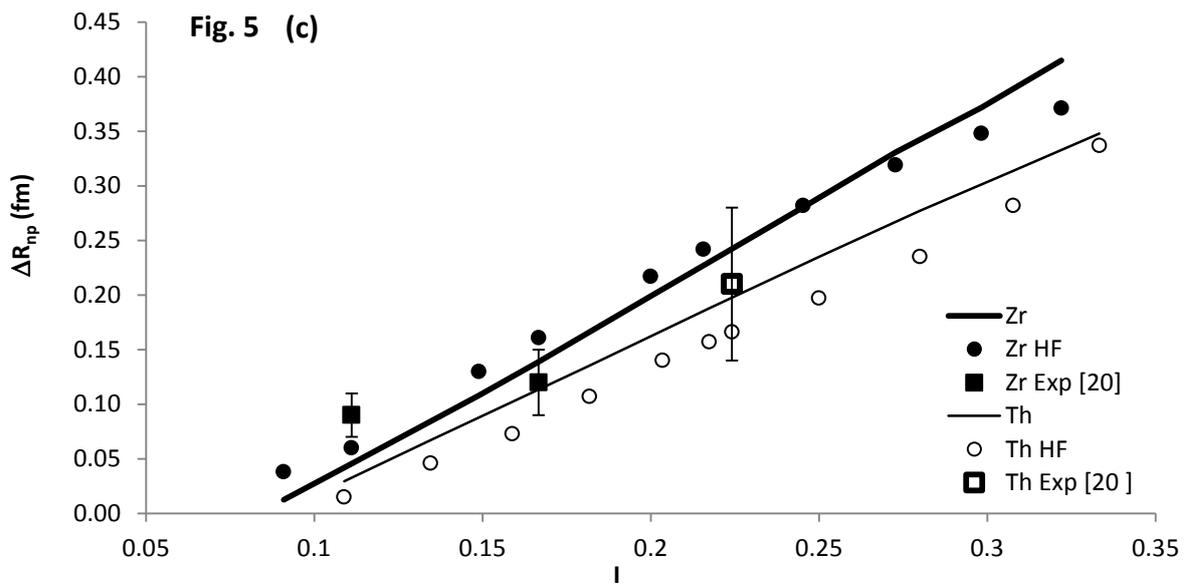

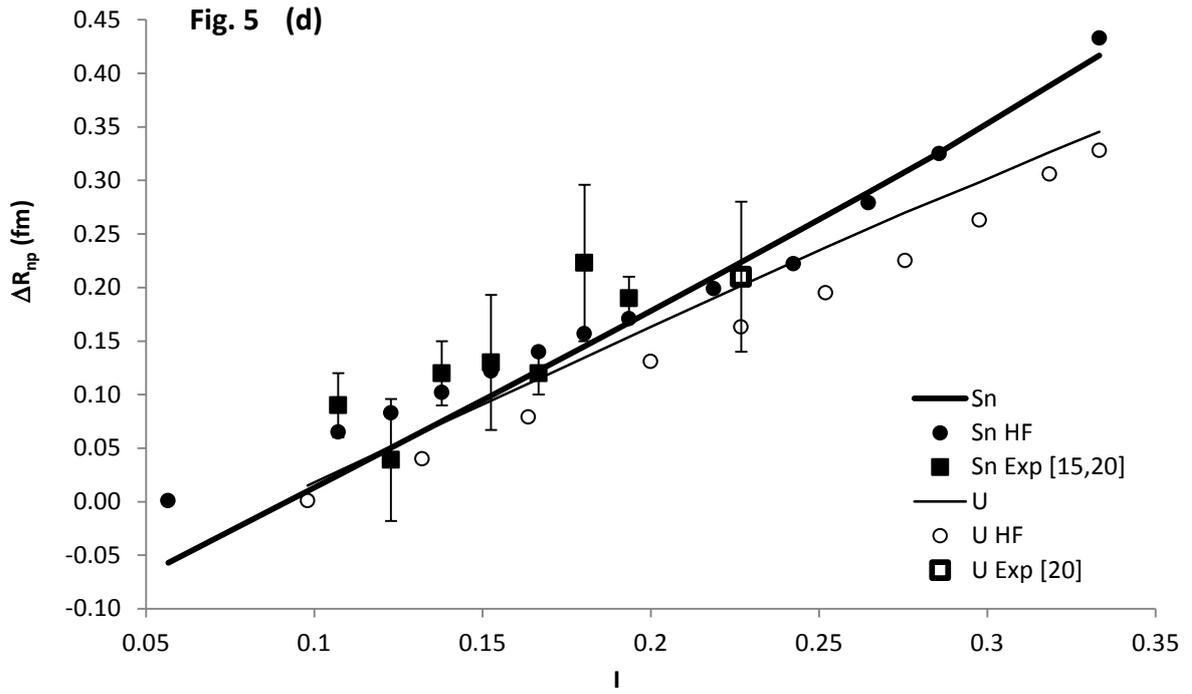

Fig. 5 (d)

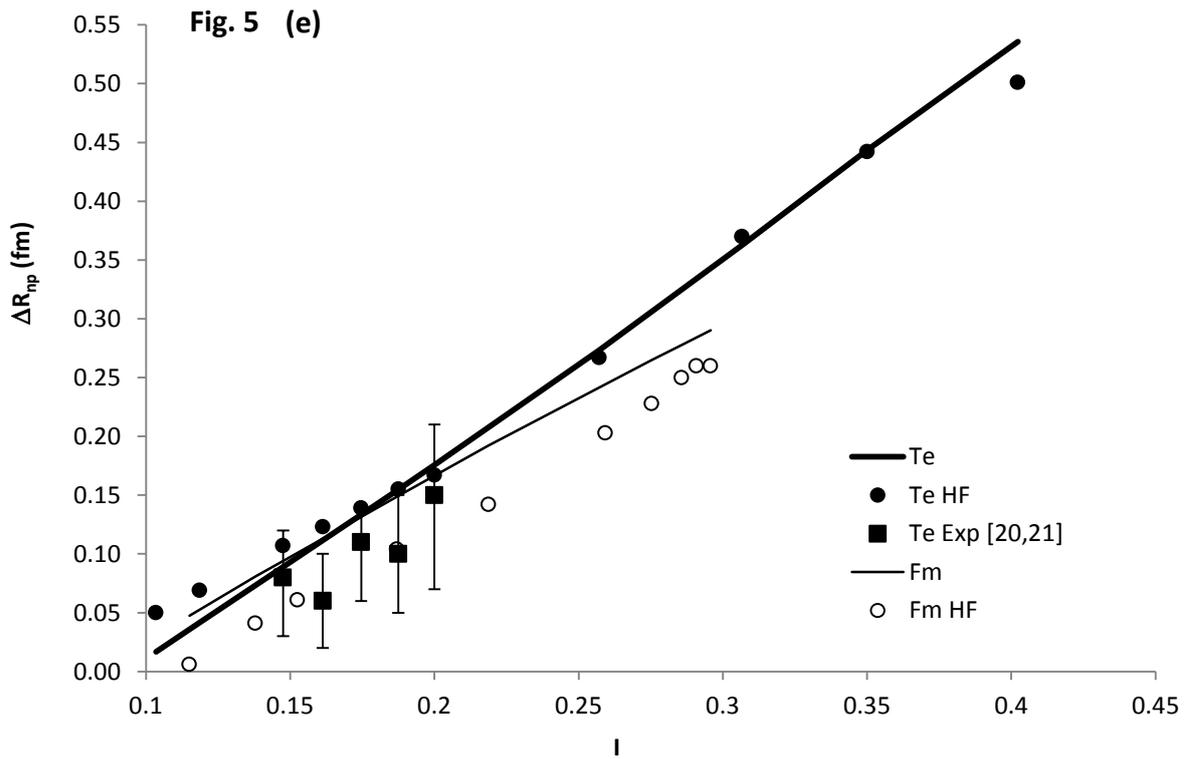

Fig. 5 (e)